\documentclass{article}

\usepackage[margin=0.75in]{geometry}
\usepackage{amsmath}
\usepackage{amssymb}
\usepackage{caption}
\usepackage{graphicx}
\usepackage{times}
\usepackage{subfigure} 
\usepackage{epstopdf}
\usepackage{float}
\usepackage{setspace}
\usepackage{hyperref}
\usepackage{url}
\begin{document}

\renewcommand\figurename{Fig.}
\renewcommand{\captionsize}{\footnotesize}
\floatstyle{plaintop}
\restylefloat{table}


\begin{center}
\LARGE \textbf{Thermal Structure and Burning Velocity of Flames in Non-volatile Fuel Suspensions} \\ [5 pt]
\large \href{http://dx.doi.org/10.1016/j.proci.2016.06.043}{\url{http://dx.doi.org/10.1016/j.proci.2016.06.043}} \\ [5 pt]

\large Michael J. Soo*, Keishi Kumashiro, Samuel Goroshin, David L. Frost, Jeffrey M. Bergthorson \\ [5 pt]
\emph{Department of Mechanical Engineering, McGill University \\ [1pt]
817 Sherbrooke Street West, Montreal, QC, H3A 0C3, Canada} \\ [5pt]
*Email: michael.soo@mail.mcgill.ca \\ [5pt]
\end{center}








\section*{Abstract}

Flame propagation through a non-volatile solid-fuel suspension is studied using a simplified, time-dependent numerical model that considers the influence of both diffusional and kinetic rates on the particle combustion process. It is assumed that particles react via a single-step, first-order Arrhenius surface reaction with an oxidizer delivered to the particle surface through gas diffusion. Unlike the majority of models previously developed for flames in suspensions, no external parameters are imposed, such as particle ignition temperature, combustion time, or the assumption of either kinetic- or diffusion-limited particle combustion regimes. Instead, it is demonstrated that these parameters are characteristic values of the flame propagation problem that must be solved together with the burning velocity, and that the \emph{a priori} imposition of these parameters from single-particle combustion data may result in erroneous predictions. It is also shown that both diffusive and kinetic reaction regimes can alternate within the same flame and that their interaction may result in non-trivial flame behavior. In fuel-lean mixtures, it is demonstrated that this interaction leads to certain particle size ranges where burning velocity increases with increasing particle size, opposite to the expected trend. For even leaner mixtures, the interplay between kinetic and diffusive reaction rates leads to the appearance of a new type of flame instability where kinetic and diffusive regimes alternate in time, resulting in a pulsating regime of flame propagation.

\section*{Keywords}
Two-phase flames, particle combustion regime, ignition, extinction, instability

\section{Introduction}

Understanding the mechanisms controlling flame propagation in suspensions of non-volatile particles is crucial to obtaining efficient combustion of metallized propellants, slurry fuels, pulverized coal, and powdered metals as carbon-free chemical energy carriers \cite{Bergthorson2015}. It is also necessary for the mitigation of catastrophic explosions in coal mines or in process industries that involve handling metallic dusts and other combustible solid powders. 

Like gas flames, flames in particulate suspensions at the laboratory scale are primarily driven by molecular heat diffusion and have comparable burning velocities \cite{Bergthorson2015}. Nevertheless, they exhibit several significant differences in their structure and behavior from homogeneous flames due to their multiphase nature.

The main distinctive feature of a flame in a solid suspension is the ability of particles to ignite--that is, to transition from a combustion regime limited by reaction kinetics to a regime limited by diffusion of the oxidizing gas towards the particle surface, or in the case of evaporating particles, towards the micro-flame enveloping each individual particle.

After ignition, the temperature of the particle or micro-flame can exceed the gas temperature by several hundred degrees, often surpassing the adiabatic flame temperature for fuel-lean mixtures. The particle combustion rate in the diffusion combustion regime is a weak, non-Arrhenius, function of gas temperature. Unlike gas flames, the width of the flame reaction zone in particle suspensions can span a large temperature range and can be comparable to, or even exceed, that of the preheat zone \cite{Goroshin1996quenching}.  The existence of diffusion micro-flames within a global flame-front (in effect, flames within the flame), which are insensitive to the bulk gas temperature, makes dust flames resistant to heat loss \cite{Bergthorson2015, Tang2011, Frank-Kamenetskii1969, Vulis1961} and also serves to maintain a constant burning velocity with increasing fuel concentration in fuel-rich mixtures \cite{Goroshin1996a}. The ability of particles to ignite, together with low ignition temperatures, may result in much wider flame propagation limits for particle suspensions than for gaseous fuels. 

Despite the overall qualitative understanding of the crucial role of the particle combustion regime on burning velocity and thermal structure, the theoretical description of flames in particulate suspensions has been limited to simple semi-empirical models that postulate either purely diffusion or purely kinetic modes of particle combustion \cite{Goroshin1996a, Seshadri1992}. The diffusive combustion models presume that particles within the suspension ignite and transit to the diffusion regime instantaneously when they reach the ignition temperature of a single, isolated particle. The common assumption is that, after ignition, the particle within the suspension will have a combustion time equal to that of an isolated single particle. Using this approach, particle ignition temperature and combustion time are considered to be external parameters that are independent of the flame-propagation mechanism. As such, they often are taken from experiments with individual particles or calculated using theoretical models for single particle combustion. These assumptions are useful for estimation, but are, in general, not justified and may lead to erroneous predictions. In reality, the particle ignition temperature, particle reaction time, and the actual regime of particle combustion are all characteristic values of the flame propagation problem directly linked to the burning velocity eigenvalue. Moreover, the particle combustion regime may alternate throughout the flame. For example, the particle may start to react in a kinetically limited regime, then transit to a diffusion-dominated combustion mode, before returning to a kinetics-dominated mode \cite{Tang2011}. As a result, a non-negligible fraction of the particle mass may be consumed during both diffusive and kinetic combustion, leading to a complex dependence of the burning velocity on particle size and concentration, as demonstrated in this study.

In this paper, the thermal structure of a flame in a particulate suspension is investigated using a simplified, transparent model that assumes that non-volatile solid fuel particles react via a single-step Arrhenius surface reaction with gaseous oxidizer delivered to the particle surface by diffusion. Besides incorporating heterogeneous reaction kinetics, the model does not impose the particle combustion mode or any other external combustion parameters. The flame propagation problem is solved numerically in a non-stationary formulation developed by Spalding \cite{Spalding1956}. This approach avoids the difficulties inherent in a steady-state formulation and permits the investigation of flame stability, which has led to the discovery of a new type of oscillating flame in heterogeneous mixtures.

\section{Model Formulation}

\subsection{Combined Kinetic-Diffusive Reaction Rate}
Following the quasi-stationary approach of Frank-Kamenetskii \cite{Frank-Kamenetskii1969}, the overall reaction rate per unit surface area of a particle in an oxidizing gas, accounting for both kinetic and diffusion ``resistances'' can be written as
\begin{equation}
\dot{\omega} = \gamma \frac{\kappa \beta}{\kappa + \beta}C_0
\end{equation}
where $\dot{\omega}$ is the particle mass consumption per unit surface area, $\gamma$ is the stoichiometric coefficient, and $C_0$ is the concentration of oxidizer in the bulk gas far from the particle surface \cite{Soo2015}. The kinetic term, $\kappa$, is the overall Arrhenius surface reaction rate, $\kappa = k_0 \exp{(-E_{\rm{a}}/R_{\rm{u}}T_{\rm{s}})}$,  where $k_0$ is the pre-exponential factor, $E_{\rm{a}}$ is the activation energy, and $R_{\rm{u}}$ is the universal gas constant. $\beta$ is the mass transfer coefficient between a particle and the gas. 

For simplicity, the contribution of Stefan flow to the heat and mass transfer between the particle and the surrounding gas is assumed to be negligible. Stefan flow is small when the molecular weight of the reaction products is close to that of the consumed oxidizer or when the initial oxidizer concentration is relatively low \cite{Frank-Kamenetskii1969}. In the absence of Stefan flow, the mass transfer coefficient takes the simple form, $\beta = \textrm{Sh} D_{\rm{i}} / 2r$. Here, $D_{\rm{i}}$ denotes the oxidizer diffusivity at the particle-gas interface, and $r$, the particle radius. For a spherical particle that is stationary relative to the gas, the exact solution to the steady-state diffusion equation yields a Sherwood number, Sh, equal to 2 \cite{Frank-Kamenetskii1969}. 

The reaction rate in Eq. (1) can then be written in terms of the normalized Damk\"{o}hler number, Da$^*$, which is related to the traditional Damk\"{o}hler number, Da $=\kappa/\beta$, by $\textrm{Da}^* = \textrm{Da}/(1 + \textrm{Da})$ \cite{Soo2015}. The resulting expression for the heat release rate per unit particle surface area is:
\begin{equation}
Q_{\rm{r}} = q \dot{\omega} = q \gamma \beta \textrm{Da}^* C_0
\end{equation}
where $q$ is the heat of reaction.

This formulation naturally incorporates the two limiting kinetic and diffusive regimes. In a kinetically controlled regime ($\beta \gg \kappa$), Da$^*$ approaches zero. In the diffusion-limited regime ($\kappa \gg \beta$), the reaction rate is primarily limited by diffusive transport of oxidizer, and Da$^*$ approaches unity.

\subsection{Ignition and Combustion of a Single Particle}
The reaction of a suspension of particles within the flame is fundamentally different from the reaction of an isolated single particle \cite{Soo2015}. However, mapping the reaction regimes for a single particle is crucial to understanding this difference and interpreting the flame structure in a suspension. The interplay between kinetic and diffusion reaction rates of heterogeneous chemical reactions leading to the processes of ignition and extinction was first investigated by Frank-Kamenetskii \cite{Frank-Kamenetskii1969} and then analyzed further by Vulis \cite{Vulis1961}. Their analyses are adapted here in a simplified form to interpret the reaction behavior of a single particle injected into a hot oxidizing gas.

The modified Semenov diagram shown in Fig. 1 plots the reaction heat release rate that accounts for both kinetic and diffusive rates as a function of the particle temperature, $T_{\rm{s}}$. At low temperatures, the reaction rate increases exponentially with $T_{\rm{s}}$ due to Arrhenius kinetics. At high temperatures, Da$^*$ approaches unity and the reaction rate in Eq. (2) becomes practically independent of temperature. If the heat loss from radiation is negligible, as is the case for moderate temperatures and small particles, then the heat loss rate is proportional to the temperature difference between the particle and gas and is plotted in Fig. 1 as straight lines: $Q_{\rm{l}} = h(T_{\rm{s}} - T_{\rm{g}})$. Here $h$ is the heat transfer coefficient, given by  $h = \textrm{Nu} \lambda_{\rm{i}} / 2r$, where $\lambda_{\rm{i}}$ is the thermal conductivity of the gas at the gas-solid interface. The Nusselt Number, Nu, is equal to 2 for a spherical particle that is stationary relative to the gas \cite{Frank-Kamenetskii1969}.

\begin{figure}[h]
\centering
\includegraphics[width=67.7mm]{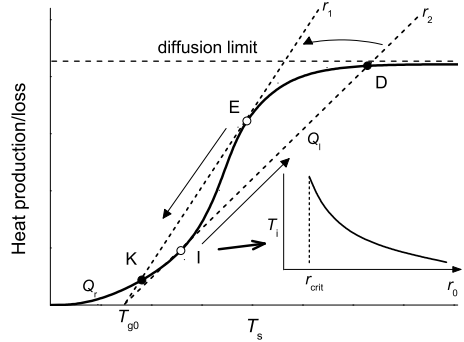}
\caption{Modified Semenov diagram illustrating particle ignition (I) at some critical temperature ($T_{\rm{g_0}}$ for $r_2$) and transition to near the diffusion limit (D). Extinction (E) occurs as the particle shrinks and undergoes the reverse transition to the kinetic limit (K). Stable states are shown by ($\bullet$) while unstable states are denoted by ($\circ$). The heat production curve is solid, and the heat loss curve is dashed. Inset: ignition temperature as a function of particle size showing critical radius below which ignition is impossible \cite{Soo2015}.}
\end{figure}

At some critical gas temperature, known as the ignition temperature, the heat loss and the heat release curves become tangent (point I on Fig. 1). The particle starts to accumulate heat and promptly transitions from a predominantly kinetic combustion regime to a predominantly diffusive regime with temperatures close to the adiabatic temperature of the stoichiometric mixture (point D). Once ignited, the burning particle inevitably extinguishes when its radius reduces to the point ($r_1$) where the heat loss and heat release curves become tangent again, albeit at a higher particle temperature (point E). After extinction, the particle will transit to a kinetic burning regime (point K). For large particles, the extinction radius is small in comparison to the initial particle size, and, thus, the residual particle mass at extinction is negligible since $m \sim r^3$. For smaller particles, the mass of the extinguished particle may not be negligible relative to the initial particle mass. If the initial particle size is reduced even further, the slope of the heat loss curve becomes so steep that tangency of the heat release and heat loss curves becomes impossible. Below this critical radius, the particle cannot ignite at any gas temperature (see inset in Fig. 1) \cite{Soo2015}.

The particle temperature history for three different initial sizes injected into a hot oxidizing gas is illustrated in Fig. 2. The first case ($r_0 \gg r_{\text{crit}}$) corresponds to a large particle that undergoes heating, then ignites and reacts to completion almost entirely in the diffusive regime. The second case ($r_0 \sim r_{\text{crit}}$) corresponds to an intermediate-sized particle that ignites but extinguishes soon afterwards without achieving full-fledged diffusive combustion. After extinction, the particle continues to react in the kinetic regime at a temperature close to the gas temperature. The third case ($r_0 < r_{\text{crit}}$) corresponds to a particle whose size is below the critical value, and, as such, reacts entirely in the kinetic regime without ignition. 

\begin{figure}[h]
\centering
\includegraphics[width=67.7mm]{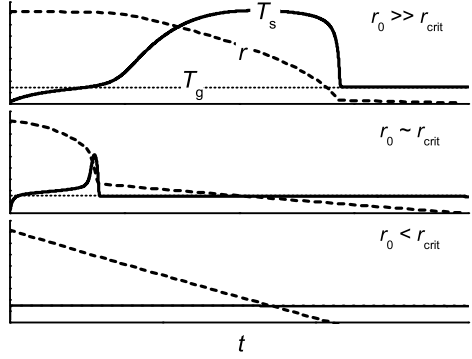}
\caption{Particle temperature and radius history of a particle injected into a hot oxdizing gas at temperature $T_{\rm{g}}$ burning in a predominantly diffusive combustion regime ($r_0 \gg r_{\text{crit}}$), an intermediate regime ($r_0 \sim r_{\text{crit}}$), and a kinetic combustion regime ($r_0 < r_{\text{crit}}$) \cite{Soo2015}.}
\end{figure}

This analysis shows that the particle reaction regime is a strong function of the particle radius as well as the surrounding temperature and oxidizer concentration. In a flame, a reacting particle with constantly changing radius is exposed to rising temperature and falling oxidizer concentration fields that are functions of the burning velocity. Therefore, postulation of the particle reaction regime and ``ignition temperature'' and ``combustion time'' parameters is, generally, unjustified in flame modelling. As demonstrated in this paper, only by solving the flame propagation problem as a whole can the particle combustion regime and other fundamental flame parameters be determined. 

\subsection{Governing Equations and Numerical Method}

The flame propagation problem is cast in a time-dependent formulation, based on the method developed by Spalding \cite{Spalding1956}, and solved numerically. This method has been used to analyze gas-phase flames \cite{Zeldovich1959}. Smoot et al. also used this method to model flames in coal dust-air mixtures \cite{DouglasSmoot1977}, but no attempt was made to draw general conclusions beyond the specifics of coal combustion.

The model presented here employs several simplifying assumptions: (i) the solid fuel can be modelled using the continuum approach described in  \cite{goroshin2011reaction,goroshin1998effect}, (ii) the velocity slip between the particles and gas is negligible, (iii) the molecular weight and heat capacity of the products and oxidizer are similar, and (iv) the solid particles do not undergo any phase transitions \cite{Zolotko1991}. All of these are second-order effects.

The governing equations for the flame are simplified by introducing a density-weighted coordinate, $x$, which is related to the physical coordinate, $x'$, by $x = \int_0^{x'} [\rho_{\rm{g}} (\bar{x'}) / \rho_{\rm{g_0}}] d\bar{x'}$ \cite{Spalding1956}. In this coordinate system, the continuity equation is automatically satisfied, and advection effects induced by thermal expansion of the gas are absent. Once found, the solution in $x$-space can be transformed into $x'$-space, as described by Margolis \cite{Margolis1978}.

For simplicity, the temperature dependences of transport and kinetic properties are chosen to eliminate thermal expansion effects. With these assumptions, the governing equations can be reduced to the following expressions for the gas-phase temperature (Eq. (3)), particle temperature (Eq. (4)), particle mass, $m_{\rm{s}}$ (Eq. (5)), and normalized oxidizer mass fraction, $Y = Y_{\rm{ox}}/Y_{\rm{ox,0}}$ (Eq. (6)). The combined kinetic-diffusion reaction rate, $\dot{\omega}$, is given by Eq. (1).
\begin{equation}
\rho_{\rm{g_0}} c_{\rm{g}} \frac{\partial T_{\rm{g}}}{\partial t} = \lambda_0 \frac{\partial ^2 T_{\rm{g}}}{\partial x^2} + N_0 A h (T_{\rm{s}} - T_{\rm{g}})
\end{equation}
\begin{equation}
c_{\rm{s}} \frac{\partial}{\partial t}(m_{\rm{s}} T_{\rm{s}}) = A q \dot{\omega} Y - A h (T_{\rm{s}} - T_{\rm{g}})
\end{equation}
\begin{equation}
\frac{\partial m_{\rm{s}}}{\partial t} = - A \dot{\omega} Y
\end{equation}
\begin{equation}
\frac{\partial Y}{\partial t} = D_0 \frac{\partial^2 Y}{\partial x^2} - \frac{N_0 \dot{\omega} A}{\gamma \rho_{\rm{{g_0}}}} Y
\end{equation}

Here, $A$ is the instantaneous particle surface area, $c_{\rm{s}}$ and $c_{\rm{g}}$ are the solid and gas-phase specific heats, $D_0$ and $\lambda_0$ are the bulk mass diffusivity and thermal conductivity of the gas, respectively, $\rho_{\rm{{g_0}}}$ is the initial gas density, and $N_0$ is the initial particle number density. The numerical values for the thermodynamic, transport and reaction properties used are given in Table 1. These values are similar to those for an aluminum particle in air, but the model is not meant to describe aluminum combustion in particular. The reaction parameters are chosen such that the switch from kinetics- to diffusion-dominated particle combustion can be observed in the micrometer particle size range. 

The equations were numerically integrated in MATLAB by the method of lines using a second-order finite difference formulation and a multi-step, variable order, implicit time integration scheme \cite{Schiesser2009}. The integration was performed over an effectively semi-infinite domain with an adiabatic and impermeable hot wall condition \cite{Margolis1978}. Because long-term solutions were of interest, the exact initial conditions used in this study turned out to be immaterial \cite{Zeldovich1985}.

\begin{table}[h]
\centering
\begin{tabular}{c c c c}
\hline
$\lambda_0$, $\lambda_{\rm{i}}$ & 0.02 W$^{1}$m$^{-1}$K$^{-1}$  & $E_{\rm{a}}/R_{\rm{u}}$ & 15 000 K  \\
$c_{\rm{g}}$ & 1010 J$^{1}$kg$^{-1}$K$^{-1}$  & $c_{\rm{s}}$ & 979 J$^{1}$kg$^{-1}$K$^{-1}$ \\
$D_0$, $D_{\rm{i}}$ & 2.0 $\times$ 10$^{-5}$ m$^2$s$^{-1}$ & $k_0$ & 70 m$^{1}$s$^{-1}$  \\
$\rho_{\rm{{g_0}}}$ & 1.17 kg$^{1}$m$^{-3}$ & $\rho_{\rm{s}}$ & 2700 kg$^{1}$m$^{-3}$\\
$\gamma$ & 1.12 & $q$ & 3.10$\times$10$^{7}$ J$^1$kg$^{-1}$\\
$C_0$ & 0.27 kg$^{1}$m$^{-3}$ & \\ \hline
\end{tabular}
\caption{Numerical values of gas and solid-fuel parameters.}
\label{tab:numVal_single}
\end{table}
 
\section{Results and Discussion}

\subsection{Flame Thermal Structure}

After an initial transient period, a steady-state flame, propagating at a constant speed, typically develops. As discussed below, for some values of $\phi$ and $r_0$, no steady-state solution is observed. Steady-state flame profiles of gas and particle temperatures, oxidizer mass fraction and particle radius are shown for the fuel-lean case ($\phi = 0.5$) in Fig. 3 for three different initial particle sizes corresponding to overall kinetic ($r_0 = 1 ~ \mathrm{\mu}$m), diffusive ($r_0 = 10 ~ \rm{\mu}$m), and intermediate ($r_0 = 4 ~ \rm{\mu}$m) particle combustion regimes.

\begin{figure*}[h]
\centering
\includegraphics[width=144mm]{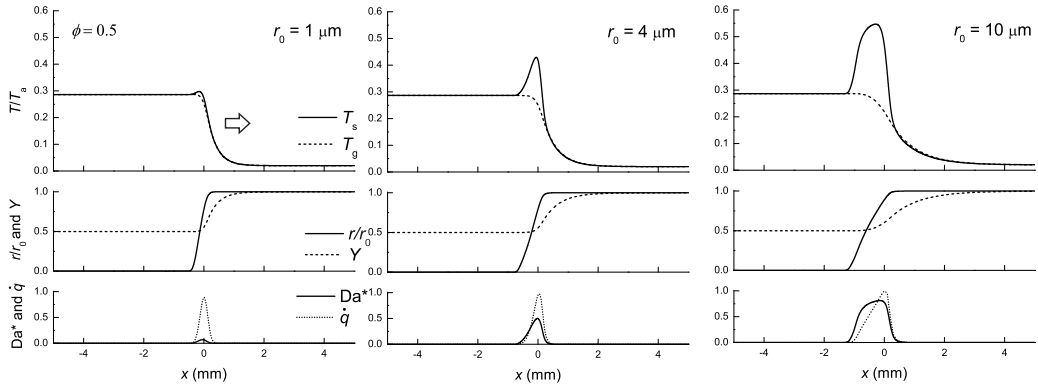}
\caption{Flame structure profiles for three different particle sizes in a fuel-lean flame ($\phi = 0.5$). Top: gas and particle temperature non-dimensionalized by $T_{\rm{a}} = E_{\rm{a}}/R_{\rm{u}}$; Middle: normalized oxidizer mass fraction, $Y$, and particle radius, $r/r_0$; Bottom: reaction heat release rate, $\dot{q}$, and Da$^*$. The arrow indicates direction of propagation.}
\end{figure*}

For small particle sizes, the flame structure is similar to that of a homogeneous gaseous flame. The gas and particle temperatures remain close to each other throughout the flame, and the reaction zone is thin and located near the point where the temperature reaches the adiabatic flame temperature of the mixture. 

In contrast, the flame structure for large particles is significantly different. After ignition, the particles attain a temperature close to the adiabatic flame temperature of the stoichiometric mixture, which greatly exceeds that of the fuel-lean mixture. In addition, the reaction zone is wide, encompassing a large gas temperature range. There is also a slight lag of the particle temperature profile behind the gas temperature in the flame preheat zone. 

In the case of intermediate particle sizes, particles ignite, but then extinguish before achieving full-fledged diffusion-limited combustion. Thus, the particle temperature surpasses the adiabatic flame temperature of the mixture, but does not reach that of the stoichiometric mixture. In contrast to the reaction of a single particle of the same size, the kinetic reaction time after the transition back to the kinetic combustion mode is relatively short due to the high flame temperatures and, thus, does not significantly increase the overall combustion time.

Flames in fuel-\emph{rich} suspensions, as shown in Fig. 4, have a different structure to that of fuel-lean mixtures, since the limiting reactant changes from the solid fuel, which has a negligible mass diffusivity, to the oxidizing gas, which readily diffuses across the flame. The primary difference is that the temperature of the particles after ignition does not exceed the adiabatic flame temperature of the mixture. The reason for this is that the temperature separation between particles and gas in the diffusive combustion regime is proportional to the concentration of the oxidizer in a particular location, which falls to zero as the temperature rises. Due to complete oxidizer consumption at the end of the combustion zone and, therefore, a negligibly small reaction rate, the value of Da$^*$ remains constant after the oxidizer is depleted. 

\begin{figure*}[h]
\centering
\includegraphics[width=144mm]{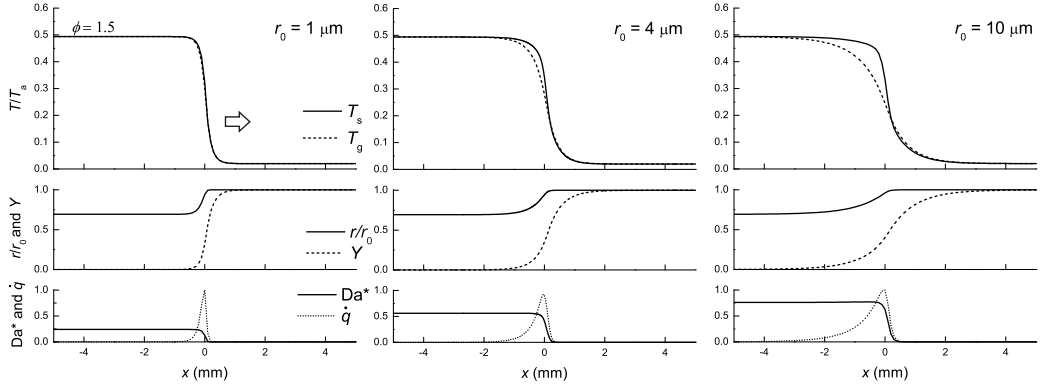}
\caption{Flame structure profiles for three different particle sizes in a fuel-rich mixture ($\phi = 1.5$). The same variables are shown as in Fig. 3.}
\end{figure*}

\subsection{Burning Velocity}

The burning velocity is calculated based on the flame-front displacement as a function of time. The burning velocity is plotted in Fig. 5 as a function of initial particle size for different equivalence ratios. The general trend for stoichiometric, rich, and somewhat lean mixtures is that the burning velocity increases monotonically with decreasing particle size. This is simply due to the increasing specific reaction surface area with decreasing particle size. In contrast, this dependence becomes non-monotonic for very lean mixtures. For equivalence ratios below $\phi = 0.4$, there is also a region where the burning velocity curves split into two separate branches between which no steady solution exists. 

\begin{figure}[h]
\centering
\includegraphics[width=67.7mm]{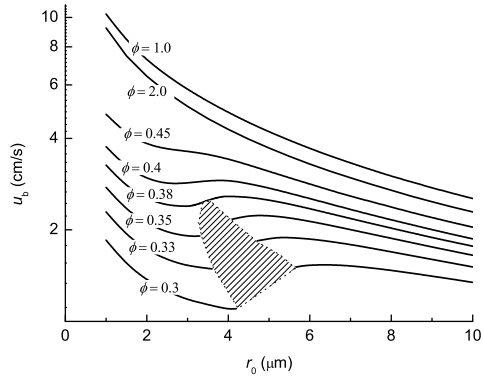}
\caption{Burning velocity, $u_{\rm{b}}$, as a function of particle size for different equivalence ratios. Hatched area corresponds to regions where unstable flame propagation is observed.}
\end{figure}

The nontrivial dependence of the burning velocity on particle size in lean mixtures is the result of the intricate interplay between the kinetic and diffusive particle combustion modes.  Because of the relatively low flame temperatures in lean mixtures, the combustion time of a larger particle reacting diffusively can be shorter than that of a smaller particle reacting kinetically. This behavior is counter-intuitive, but is a direct consequence of the ability of large particles to ignite and burn with temperatures much higher than the adiabatic flame temperature of the lean mixture. The appearance of ``flames within the flame", results in higher reaction rates for suspensions of larger particles that offset, and potentially even overcome, the reduced specific reaction surface area as compared to finer suspensions. This behavior can be seen first as a plateau in the dependence of burning velocity on particle radius (Fig. 5, $\phi = 0.4$) and then a decline as particle size decreases (Fig. 5, $\phi < 0.4$).  It can also be seen in Fig. 6 as intersections of the burning velocity versus equivalence ratio curves for different particle sizes. As the particle size is increased, eventually the effect of decreasing specific reaction surface area again dominates, and the burning velocity decreases with increasing particle size. These trends have been observed experimentally in the combustion of coal suspensions \cite{DouglasSmoot1977} and liquid sprays \cite{Chan1988}, although no clear physical explanation was given at the time. 

\begin{figure}[h]
\centering
\includegraphics[width=67.7mm]{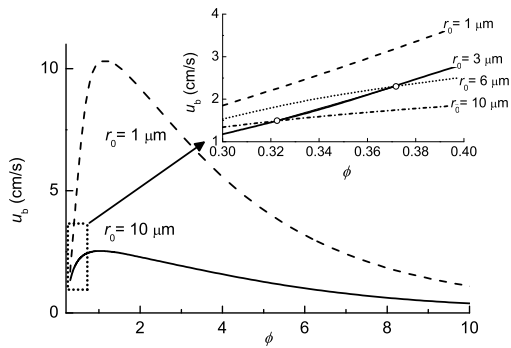}
\caption{Burning velocity as a function of equivalence ratio for different initial particle sizes. The expanded view shows the intersections of burning velocities for lean equivalence ratios for various particle sizes.}
\end{figure}

\subsection{Kinetic-Diffusive Instabilities}

For equivalence ratios below $\phi = 0.4$, there is a region where no steady-state flame propagation occurs (see hatched area of Fig. 5). Instead, these flames oscillate both in terms of burning velocity and flame structure. An example of instantaneous profiles of particle and gas temperature, and Da$^*$ at different instances over a single oscillation period are shown in Fig. 7 for an equivalence ratio of $\phi = 0.4$ and initial particle size of $r_0 = 4~\rm{\mu}$m. 

\begin{figure}[h]
\centering
\includegraphics[width=67.7mm]{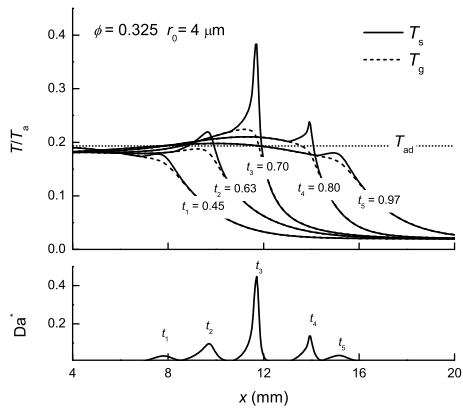}
\caption{Instantenous profiles of gas and particle temperatures and Da$^*$ at different instances in time over a single oscillation period of an oscillating flame. The units of $t$ are seconds. The arrow indicates the direction of propagation.}
\end{figure}

The gas temperature is observed to oscillate about the adiabatic flame temperature, while the particle temperature oscillates with a much greater amplitude than that of the gas. At $t_3$, the combustion regime is more diffusive, as indicated by the large temperature separation and Da$^*$, while at $t_1$ and $t_5$, the regime is more kinetic, as indicated by the low values of temperature separation and Da$^*$.

At first glance, these oscillations resemble the well-known thermo-diffusive flame instability often observed in homogeneous gas \cite{Christiansen1998} and quasi-homogeneous solid mixtures \cite{Yi1990}, and also recently observed in aluminum dust flames \cite{Julien2015}. The physical mechanisms underlying the thermo-diffusive instability and the heterogeneous pulsating flame shown here are, however, fundamentally different. In homogeneous mixtures, pulsations are induced by changes in the gas molecular transport properties (Lewis number), activation energy, and overall heat production, as required by the thermo-diffusive instability criteria formulated by Matkowsky and coworkers \cite{Matkowsky1978, Matkowsky1980}. In the particle flame, the fuel has zero diffusivity, and, for a fixed equivalence ratio, the gas-phase transport, kinetic parameters, and overall heat production are fixed. In particular, they do not change with particle size. Instead, the cause of the observed instability is evidently rooted in the processes of particle ignition and extinction that are, indeed, very sensitive to particle size. Thus, it may be called a kinetic-diffusive instability, reflecting the fact that the particle combustion regime in the pulsating flame oscillates between diffusive and kinetic combustion modes. This interesting new phenomenon will be studied in detail in subsequent work.

\subsection{Comparison to Semi-Empirical Flame Models}

Most existing models for flames in suspensions are semi-empirical and postulate either a purely kinetic or purely diffusive regime of particle combustion. Plots of the normalized burning velocity versus particle size predicted by such models are compared with the result of the current numerical simulations in Fig. 8 for the lean mixture with an equivalence ratio of $\phi = 0.3$. 

\begin{figure}[h]
\centering
\includegraphics[width=67.7mm]{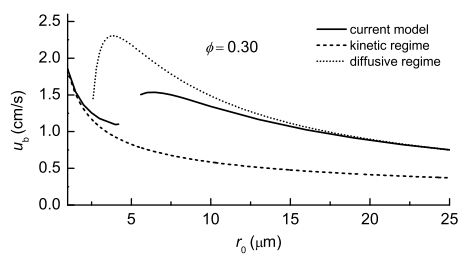}
\caption{Comparison of computed burning velocity to semi-empirical asymptotic Zeldovich and Frank-Kamenetskii (kinetic) and modified Mallard-Le Chatelier (diffusive) flame models. The semi-empirical models are normalized to the numerical model at $r_0 = 1~\rm{\mu}$m for the kinetic model and $r_0 = 25~\rm{\mu}$m for the diffusive model.}
\end{figure}

For the case of the purely kinetic regime, the burning velocity is calculated using an analytical expression formulated by Zeldovich and Frank-Kamenetskii, modified for a combustible mixture with a single-step, first order heterogeneous surface reaction \cite{Zeldovich1985}. The result is normalized to match that of numerical simulations for the smallest particles, which burn in the kinetic regime. 

The burning velocity in the case of the purely diffusive regime is calculated using an expression given in \cite{Goroshin1996quenching, Goroshin1996a} that requires two externally defined combustion parameters: the ignition temperature and combustion time of the single particle. Ignition temperatures at different particle sizes are calculated using the Semenov ignition criterion as shown in Fig. 1 and have a dependence with particle size similar to that shown in the inset of Fig. 1. The particle combustion time is calculated using a simple diffusive reaction model that predicts a $t_{\rm{c}} \sim r_0^2$ dependence \cite{Tang2009}. The result of the purely diffusive model is normalized to match that of the numerical simulations for the largest particles, which burn in the diffusive regime.

As can be seen in Fig. 8, these semi-empirical models only qualitatively predict the asymptotic behavior of burning velocity on particle size where the combustion regimes are either purely kinetic or diffusive. Over a wide particle size range, where diffusion and kinetic rates are comparable, these models cannot capture the physics of the flame propagation and yield erroneous results.

\section{Conclusion}

The comprehensive analysis of flame propagation in non-volatile particle suspensions presented in this paper has a historical parallel. It can be compared with the replacement of the semi-empirical Mallard-Le Chatelier flame model, based on the notion of ignition temperature and reaction time as external parameters, to flame models developed in the 20th century based on the Arrhenius reaction law. Here, it is similarly shown that the particle ignition temperature, particle reaction time, and regime of particle combustion are, in general, characteristic values of the flame problem that must be solved for together with the burning velocity. Moreover, this analysis demonstrates that the thermal structure of flames in suspensions is defined primarily by the interplay between kinetic and diffusive reaction modes that also yields non-trivial dependence of the burning velocity on particle size. This interplay leads to the emergence of a new type of pulsating instability that is related to the process of ignition and extinction of particles within the flame.

\section*{Acknowledgments}

Support for this work was provided by the Defense Threat Reduction Agency under contract HDTRA1-11- 1-0014, the NSERC CREATE on Clean Combustion Engines, and a McGill Engineering Doctoral Award. Additional funding was provided by the Panda Faculty Scholarship in Sustainable Engineering \& Design and a William Dawson Scholarship.

\let\itshape\upshape
\bibliographystyle{pci}
\bibliography{manuscript.bib}

\newpage



\newpage

\end{document}